\documentclass[aps,prb,twocolumn,superscriptaddress,amsmath]{revtex4-2}
\usepackage{graphicx}
\usepackage{hyperref}
\usepackage{mathrsfs}
\usepackage{bm}
\usepackage{color}
\usepackage{siunitx}
\hypersetup{hypertex=true,
	colorlinks=true,
	anchorcolor=blue,
	linkcolor=blue,
    citecolor=blue}
\begin{document}

\title{Spin phase regulated spin Josephson supercurrent in topological superconductor}

\author{Yue Mao}
\affiliation{International Center for Quantum Materials, School of Physics, Peking University, Beijing 100871, China}
\author{Qing-Feng Sun}
\email[]{sunqf@pku.edu.cn}
\affiliation{International Center for Quantum Materials, School of Physics, Peking University, Beijing 100871, China}
\affiliation{CAS Center for Excellence in Topological Quantum Computation, University of Chinese Academy of Sciences, Beijing 100190, China}

\date{\today}

\begin{abstract}
Without applied bias voltage,
a superconducting phase difference can drive a charge Josephson supercurrent in a superconductor junction.
In analogy, we here theoretically propose a spin phase that intrinsically generates spin Josephson supercurrent,
and this spin Josephson effect is studied in a junction of superconducting nanowire (SNW).
We show that spin-orbit coupling and magnetic field give rise to spin-triplet superconductivity in the SNW,
thus allow the superfluid of both charge and spin.
Next we introduce the concept of spin phase that
can be generated by controls of spin current, magnetic field or electric field.
By the analysis of pairing correlations and Ginzburg-Landau-type theory,
it is shown that the spin phase makes spin-up and spin-down $S=1$ Cooper pairs get opposite phases
and move oppositely in a Josephson junction,
so that a dissipationless pure spin current is induced.
We also derive the formula
that the spin current is equal to the derivative of
Andreev bound state to the spin phase,
which is analogous to that of charge Josephson effect.
At last, our calculation verifies the existence of spin supercurrent
inside the superconducting gap
in the topologically nontrivial phase.
Our study provides a view on spin Josephson effect and indicates
the potential combination of topological superconductivity and spintronics.
\end{abstract}

\maketitle

\section{\label{SEC1} Introduction}

One of milestones in superconductivity is Josephson effect \cite{Josephson1962}:
for a junction of two superconductors,
in the presence of a superconducting phase difference,
the tunneling of Cooper pairs can induce a supercurrent
without applied voltage bias \cite{Bagwell1992_Jose,Golubov2004_Jose}.
For conventional superconductors, the Cooper pairs are spin-singlet,
so the junction has only charge current but no spin current.
From the perspective of spin, the Josephson effect has analogies in spin superfluid, such as spin-triplet exciton systems \cite{Sun2011_SSC} and especially magnon systems \cite{Loss2014_Magnon,Loss2015_Magnon}, where a spin current can also be generated by a phase bias \cite{H2016_Magnon}.
The carriers of these systems are bosonic spin excitations that can condense into superfluid and form a spin supercurrent \cite{Sun2011_SSC,Sun2013_SSC,Spintronics2015_Magnon,Duan2018_SSC}.
Searching for this dissipationless spin current is an intriguing subject because of its outstanding performance in reducing energy consumption and storing information \cite{Fert2008_Spintronics,Eschrig2015_SJE}.

A more direct idea for spin Josephson supercurrent is to use spin-triplet superconductivity \cite{Eschrig2015_SJE,Xiao2006_STSC,Birge2010_STSC,Blamire2018_STSC,Mackenzie2003_STSC}.
Although the intrinsic spin-triplet superconducting materials are rare,
this superconductivity can be obtained by constructing normal superconductor-ferromagnet heterostructures \cite{Grein2009_SJE2,Linder2010_SJE1,Belzig2011_SJE4,Hikino2013_Jose,Linder2016_SJE5}.
The proximity effect combines magnetism and superconductivity, leading to spin-triplet order parameter near the interface \cite{Eschrig2015_SJE,Xiao2006_STSC,Blamire2018_STSC,Eschrig2018_SJE7,Linder2015_SJE3}.
There are many proposals to realize the spin Josephson effect, such as operating spin-orbit coupling \cite{Barash2004_SJE6} and magnetization direction \cite{Linder2010_SJE1,Belzig2011_SJE4,Hikino2013_Jose,Linder2015_SJE3}.
In effect, these methods can induce a spin Josephson current.
Behind these methods, there could exist a spin phase that explains the spin Josephson effect, like the charge Josephson effect driven by a superconducting phase.

In this paper, we propose the concept of spin phase and report a spin Josephson supercurrent in topological superconductor.
We study spin Josephson effect of a 1D semiconductor nanowire under a magnetic field and in proximity to an $s$-wave superconductor.
This superconducting nanowire (SNW) was initially studied because of the topological phase transition and Majorana zero modes \cite{Elliott2015_MZMRMP,Lutchyn2010_MZMT,Oreg2010_MZMT,Mourik2012_MZME}, with both charge and spin conductance inside the gap \cite{Law2009_ZBP,He2014_AR,Mao2021_Spinbias}.
This can be viewed as a result of $p$-wave superconductivity \cite{Kitaev2001_MZM}.
Because of the spin-orbit coupling (SOC) and the magnetic field, spin is not a good quantum number, and there exists a spin-triplet superconducting component.
Therefore, this SNW can be managed to realize a spin Josephson effect.
In analogy to the (charge) phase that leads to the transport of charge \cite{Asano2006_Jose,Brydon2009_Jose}, we introduce the spin phase that intrinsically drives the motion of spin.
Spin is a vector, and so is spin phase.
The spin phase of SNW can be caused by controlling spin current, magnetic field or electric field.
The effect of spin phase is analyzed by pairing correlations and Ginzburg-Landau type (GL-type) theory \cite{Bao2015_SSCGL1,Lv2017_SSCGL2}.
We find that spin phase induces opposite phases on $S=1$ Cooper pairs with opposite spins, thus a pure spin supercurrent can be generated in a Josephson junction.
At last, we calculate the subgap spin current carried by Andreev bound states (ABSs), confirming the existence of the spin Josephson supercurrent.
This subgap spin current exists in topologically nontrivial phase, but is absent in the trivial phase.

The rest of this paper is as follows:
In Sec. \ref{SEC2}, we show spin-triplet superconductivity of the SNW.
In Sec. \ref{SEC3}, we introduce the concept of spin phase, discuss how to generate spin phase, and analyze its effect.
In Sec. \ref{SEC4}, we show the results of ABSs and subgap spin supercurrents.
Discussion and conclusion are given in Sec. \ref{SEC5}.
Appendix \ref{AA} and \ref{AB} give the detailed calculations and formula derivations for currents and ABSs.

\section{\label{SEC2} Spin-triplet superconductivity}

In the basis $(\psi_z, \psi_{\bar z})^T$,
we can write the Hamiltonian of a semiconductor nanowire under magnetic field as
\begin{equation}
	H_1 (p_x)= \frac{p_x^2}{2m}-\mu+\alpha\sigma_3 p_x+B\sigma_1, \label{H1}
\end{equation}
where $\sigma$'s are Pauli matrices in spin space, $\psi_{z (\bar z)}$ annihilates a spin-$z$ ($-z$) electron, $m$ is the effective mass of electrons, $\mu$ indicates the chemical potential,
$B=\frac{1}{2}g\mu_B B_{\bar x}$ is the Zeeman splitting energy caused by an $\bar x$-direction magnetic field $B_{\bar x}$, with the Bohr magneton $\mu_B$ and Land\'e $g$ factor absorbed.
The SOC $\alpha$ term originates from the built-in electric field: $ H_{\rm SO}={\bm \alpha}\cdot{\bm \sigma}p_x \propto ({\bm \sigma}\times {\bm E}_{\rm SO})\cdot {\bm p} $ \cite{Molenkamp2001_Rashba,Sun2005_Rashba}.
In the 1D system along $x$ axis, the momentum $ {\bm p} $ has only $p_x$ component, so ${\bm \alpha}$ is in the $yOz$ plane.
Here we assume a $y$-direction electric field that relates to the $\alpha=\alpha_z$ term.
In proximity to an $s$-wave superconductor, the system becomes a SNW and can be described by the mean field Hamiltonian in the basis $(\psi_z, \psi_{\bar z}, \psi_z^\dagger, {\psi_{\bar z}}^\dagger)^T$ \cite{Lutchyn2010_MZMT,Oreg2010_MZMT,Yan2019_BdG,Tkachov2015_TIBOOK,Law2016_d_vector},
\begin{equation}
	H_{\rm BdG} (p_x)=
	\begin{pmatrix}
		H_1 (p_x) & i\Delta\sigma_2 \\
		-i\Delta\sigma_2 & -H_1 ^*(-p_x) \\
	\end{pmatrix} ,\label{H_SNW}
\end{equation}\\
with $\Delta$ the induced pairing potential.
We concern the physics inside the induced gap, which is also inside the original bulk superconducting gap \cite{Lutchyn2018_MZME}, and this description for proximity effect can work well \cite{Aguado2021_delta}.

For a SNW of infinite length, $ p_x $ can be replaced by a good quantum number $ \hbar k $, with $k$ the wave vector.
The pairing symmetry at energy $\epsilon$ is described by pairing correlations \cite{Balian1963_d_vector,Tkachov2015_TIBOOK,Law2016_d_vector},
\begin{equation}
 F_{\beta \gamma}(\epsilon) = -i \int_{0}^{+\infty} dt e^{i(\epsilon+i0^+) t} \langle \{ {\psi}_{\beta}(t), {\psi}_{\gamma}(0)  \} \rangle . \label{Corre}
\end{equation}
The $2 \times 2$ matrix form of pairing correlations can be represented by $ {\bm d} $-vector ${\bm F}_{2 \times 2}= (d_0+{\bm d}\cdot{\bm \sigma})i\sigma_2 $, where $ d_0 $ represents the traditional spin-singlet pairing and $ \bm d $ denotes the spin-triplet pairing of different spin directions \cite{Balian1963_d_vector,Tkachov2015_TIBOOK,Law2016_d_vector}.
We find
\begin{equation}
	{\bm d}=\frac{2\Delta}{M}(B\epsilon,iAB,A\xi), \label{d}
\end{equation}
where $M=(B^2+\epsilon^2-\Delta^2-\xi^2-A^2 )^2+4(A^2 B^2-A^2 \xi^2-B^2 \epsilon^2)$, $ \xi=\frac{\hbar^2 k^2}{2m}-\mu $, and $ A=\alpha \hbar k $.
One can see that the SOC and magnetic field effectively leads to spin-triplet correlations, although the proximity of superconductor only causes a spin-singlet pairing potential.

\begin{figure}
	\includegraphics[width=1\columnwidth]{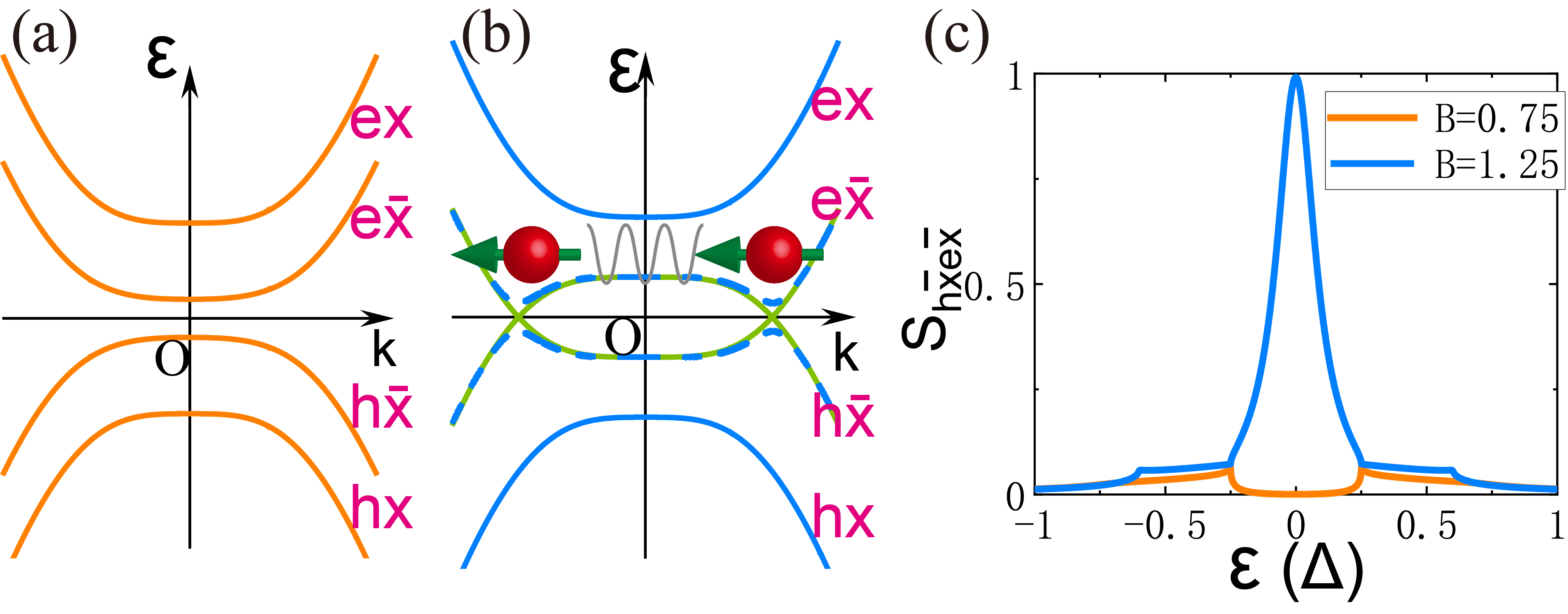}%
	\caption{\label{FIG1} Schematic plots of energy bands for (a) topologically trivial phase and (b) nontrivial phase.
	$ex$ ($e \bar x$) and $hx$ ($h \bar x$) approximately indicate electron and hole bands with $x$ ($-x$) spin.
	In (b), a high magnetic field gives rise to the overlap between $e \bar x$ and $h \bar x$, and their gap is opened by the SOC.
	Therefore, a Cooper pair is formed by two electrons with the same spin $\bar x$, and the spin-triplet component is greatly enhanced.
	(c) shows the equal spin Andreev reflection coefficient $S_{h \bar x e \bar x}$
	of the NNW-SNW junction versus the incident energy $\epsilon$.
	Here we set $\Delta, \hbar, m=1$ as units, $ \mu=0 $, and $ \alpha=0.6 $ adapted from the experiment \cite{Mourik2012_MZME}.
	The chemical potential of NNW is $ \mu_{\rm NNW}=50 $.
	Without further explanation, the parameters are the same in other figures.
	}
\end{figure}

In addition, the spin-triplet component can be enhanced by the topological phase transition, which happens at $B^2=\Delta^2+\mu^2$ \cite{Lutchyn2010_MZMT,Oreg2010_MZMT}.
In the topologically trivial phase ($B^2<\Delta^2+\mu^2$), there is a $\Delta$-dominated gap \cite{Oreg2010_MZMT}, which looks like a couple of $s$-wave bands shifted by the Zeeman effect (Fig. \ref{FIG1}(a)).
The spin-triplet pairing already exists but is not strong.
By increasing the magnetic field, the system is regulated to the nontrivial phase ($B^2>\Delta^2+\mu^2$).
In this case, the energy gap is dominated by the $B$ term \cite{Oreg2010_MZMT}.
The magnetic field strongly splits the spin-$x$ and spin-$\bar x$ bands, making an overlap between spin-$\bar x$ electron and spin-$\bar x$ hole.
The electron and hole with $\bar x$ spin can be effectively coupled by the pairing potential and the SOC, opening a distinct gap (Fig. \ref{FIG1}(b)).
This gap is the result of coupling between electrons and holes with the same spin,
so it implies a \emph{huge spin-triplet superconductivity}, with spin polarized nearly \emph{parallel to the magnetic field}.

We can see this by calculating the equal spin
Andreev reflection coefficient $S_{h \bar x e \bar x}$ of
an incident spin-$\bar x$ electron in the normal
nanowire (NNW)-SNW junction (Fig. \ref{FIG1}(c)) \cite{He2014_AR,Mao2021_Spinbias}.
There is no SOC or pairing potential in the NNW.
The Zeeman energy in NNW is neglected because its Land\'e $g$ factor is much lower than that of SNW \cite{Mourik2012_MZME}.
This Andreev reflection process is equivalent to transmitting a spin-$\bar x$-polarized Cooper pair into the SNW, which reflects the superconductivity for both charge and spin.
For the topologically trivial phase ($B=0.75$), $S_{h \bar x e \bar x}$ shows the shape of a hard gap that maximizes at band edges.
By regulating the SNW to the nontrivial phase ($B=1.25$), the coefficient is greatly increased, especially inside the gap.
It shows a peak and almost reaches 1 at $\epsilon=0$, like the conductance spectrum of a $p$-wave superconductor \cite{Asano2007_pwave,Sarma2015_Conductance}.
Therefore, the nontrivial phase of SNW indeed has a strong spin-triplet superconductivity with $S=1$ Cooper pairs nearly polarized at $\bar x$ direction.

\section{\label{SEC3} Spin phase}

Based on the spin-triplet superconductivity of SNW, a spin phase difference could be managed to drive a spin supercurrent in a Josephson junction.

The traditional charge Josephson effect can be caused by applying a charge current to a Josephson junction, because the lifted free energy results in a charge phase difference.
Formally, the charge phase difference is equivalent to a local (e.g. only the right side) unitary transformation $ \psi_{\rm R \uparrow (\downarrow)}' = \psi_{\rm R \uparrow (\downarrow)} e^{-i\phi_c/2} $.
Because the U(1) symmetry is broken, this transformation induces a phase in electron-hole coupling term \cite{Josephson1962,Golubov2004_Jose,Bagwell1992_Jose}, i.e. the pairing term is written as $\Delta_{\rm R} \psi_{\rm R \downarrow} \psi_{\rm R \uparrow} = \Delta_{\rm R} e^{i\phi_c} \psi_{\rm R \downarrow}' \psi_{\rm R \uparrow}' $.
Meanwhile, all the other terms keep unchanged.

\begin{figure}
	\includegraphics[width=\columnwidth]{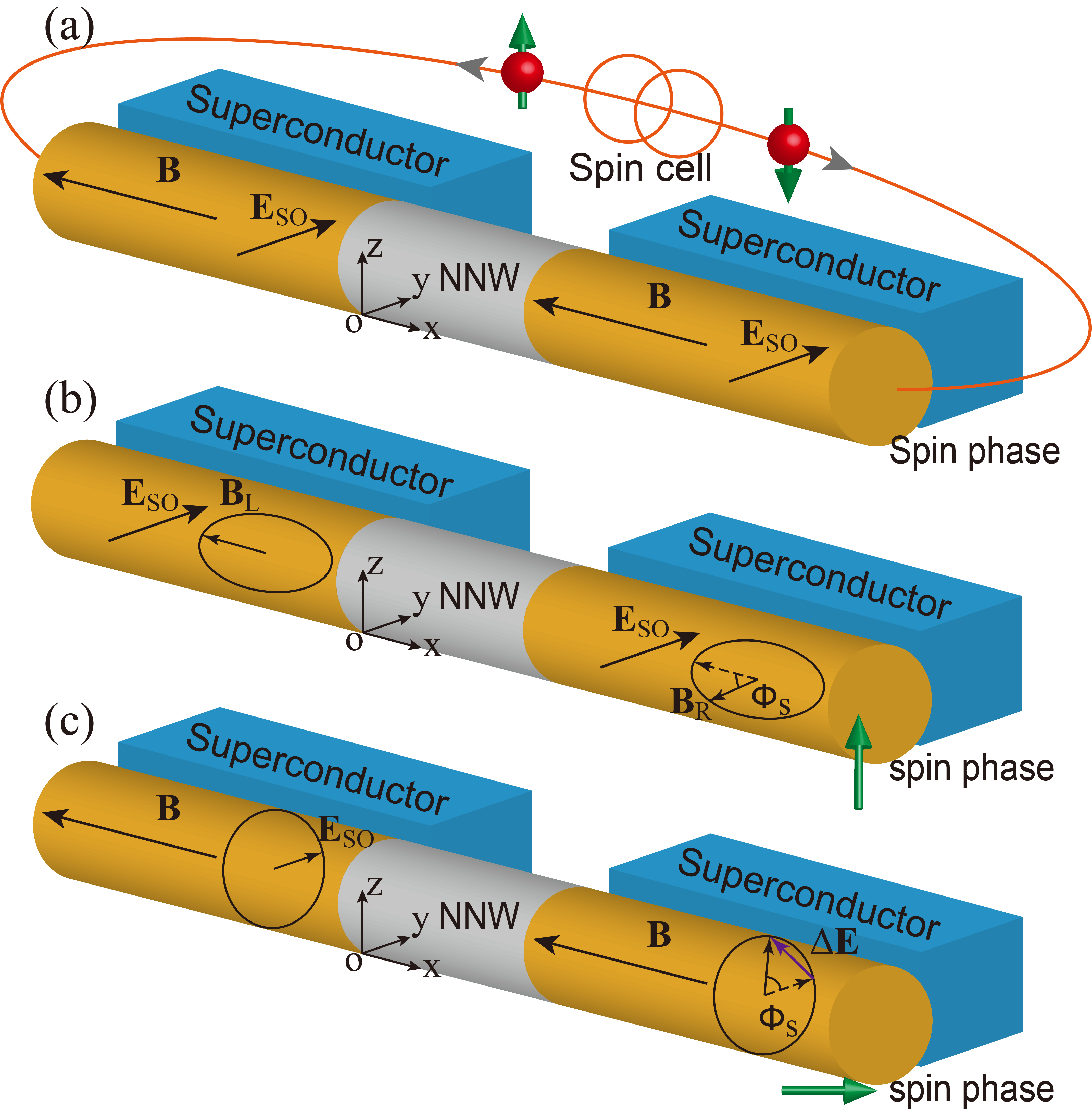}%
	\caption{\label{FIG2} The devices to induce spin phase differences.
	(a) Applying a spin current gives rise to a spin phase difference parallel
	to spin direction of the spin current.
	(b) and (c) The spin phase differences of direction $z$ and $x$ can also be caused by rotating the magnetic field (b) and regulating the electric field (c) on the right side, respectively.
	}
\end{figure}

Correspondingly, applying a spin current can generate a spin phase difference, as shown in Fig. \ref{FIG2}(a).
Because spin is a vector, spin phase is also a vector.
The direction of spin phase corresponds to the spin direction of the spin current.
By controlling the spin cell, spin phases of $x, y, z$ directions can all be obtained.
A pure $n$-direction spin current in electron systems is the result of spin-$n$ and spin-$\bar n$ electrons moving in opposite directions.
So an $n$-direction spin phase difference should equal to a couple of local unitary transformations of spin-$n$ and spin-$\bar n$ annihilating operators $ \psi_{\rm R n}' = \psi_{\rm R n} e^{-i\phi_s^n/2} $, $ \psi_{\rm R \bar n}' = \psi_{\rm R \bar n} e^{i\phi_s^n/2} $.
As a result, a phase $e^{i \phi_s^n}$ appears in $ \psi_{\rm R n} \psi_{\rm R n}$ and $ \psi_{\rm R \bar n}^{\dagger} \psi_{\rm R n} $ terms, meanwhile an opposite phase $e^{-i \phi_s^n}$ appears in $ \psi_{\rm R \bar n} \psi_{\rm R \bar n}$ and $\psi_{\rm R n}^{\dagger} \psi_{\rm R \bar n} $ terms.
The $ \psi_{\rm R n} \psi_{\rm R n}$ and $ \psi_{\rm R \bar n} \psi_{\rm R \bar n}$
terms are direct spin-triplet pairing potentials,
which can exist in $p$-wave superconductor \cite{Mackenzie2003_STSC},
but are not included in the SNW system.
The other two terms relate to SOC and Zeeman splitting.
According to Eq. (\ref{d}), these terms are origins of spin-triplet correlations,
thus the control on them can also induce a spin phase on Cooper pairs equivalently.

To get the Hamiltonian with a spin phase, we just need to operate on Hamiltonian Eq. (\ref{H1}).
Its complete BdG form is obtained by replacing $H_1 (p_x)$ in Eq. (\ref{H_SNW}) by the operated Hamiltonian.
For a $z$-direction spin phase, the corresponding terms are Zeeman splitting $ B \psi_{\rm R z}^{\dagger} \psi_{\rm R \bar z} $ and $ B \psi_{\rm R \bar z}^{\dagger} \psi_{\rm R z}$ that originate from the $\bar x$-direction magnetic field.
Applying a spin phase means that these terms become $ B e^{-i \phi_s^z} \psi_{\rm R z}^{\dagger} \psi_{\rm R \bar z} $ and $ B e^{i \phi_s^z} \psi_{\rm R \bar z}^{\dagger} \psi_{\rm R z} $, i.e.
\begin{equation}
	H_2 (p_x)= \frac{p_x^2}{2m}-\mu+\alpha\sigma_3 p_x+B(\sigma_1 \cos{\phi_s^z} + \sigma_2 \sin{\phi_s^z}). \label{H2}
\end{equation}
So, besides applying a spin current, the $z$-direction spin
phase can also be realized by rotating the magnetic field around $z$-axis by an angle $\phi_s^z$, as shown in Fig. \ref{FIG2}(b).
To induce the spin phase at $x$-direction, we should operate $ -i \alpha p_x \psi_{\rm R x}^{\dagger} \psi_{\rm R \bar x} $ and $ i \alpha p_x \psi_{\rm R \bar x}^{\dagger} \psi_{\rm R x}$ terms, i.e.
\begin{equation}
	H_3 (p_x)= \frac{p_x^2}{2m}-\mu+\alpha(\sigma_3 \cos{\phi_s^x} - \sigma_2 \sin{\phi_s^x} ) p_x+B\sigma_1. \label{H3}
\end{equation}
This can also be reached by rotating the $\bm \alpha$ term around $x$ axis.
Note that the control on SOC is intrinsically to rotate the electric field \cite{Sun2005_Rashba}.
The SNW naturally has a $y$-direction built-in electric field, and its rotation can be realized by adding an artificial field shown in Fig. \ref{FIG2}(c).

Let us discuss the spin phase in four ways:

(a) From the perspective of pairing correlations:
For charge Josephson effect, the charge phase $ \phi_c $ appears in the whole correlation matrix $ {\bm F}'_{2 \times 2}={\bm F}_{2 \times 2} e^{-i \phi_c} $.
In a Josephson junction, the Cooper pairs with any spin are equally driven by $ \phi_c $, and they move in the same direction.
This charge phase drives only charge current in conventional superconductors.
For spin-triplet superconductors, the spin-polarized Cooper pairs allow spin in a certain direction to flow with charge as an accessory \cite{Asano2006_Jose,Brydon2009_Jose}.
For our spin Josephson effect, the initial pairing correlations of SNW are $ F_{zz}=-\frac{2\Delta B(A+\epsilon)}{M}$ and $F_{\bar z \bar z}=-\frac{2\Delta B(A-\epsilon)}{M} $.
After adding a $z$-direction spin phase, the new correlations are $ F'_{zz} =F_{zz} e^{-i\phi_s^z} $ and $ F'_{\bar z \bar z} = F_{\bar z \bar z} e^{i\phi_s^z} $, while the other two terms $F_{z \bar z}$ and $ F_{\bar z z} $ are unchanged.
As for adding an $x$-direction spin phase, we can also find that the nonzero correlations become $F'_{xx} = F_{xx} e^{-i\phi_s^x} $ and $ F'_{\bar x \bar x} = F_{\bar x \bar x} e^{i\phi_s^x} $.
This means that Cooper pairs with $z$ ($x$) and $\bar z$ ($ \bar x $) spins get \emph{opposite phases}.
So they move in opposite directions and a \emph{spin Josephson supercurrent} is induced.
In general, for adding an arbitrary $n$-direction spin
phase with $\hat n=(\sin \theta \cos \phi, \sin \theta \sin \phi, \cos \theta)^T$, we have
\begin{eqnarray}
	\bm F'_n & \equiv & \begin{pmatrix}
	F'_{nn} & F'_{n\bar{n}} \\
	 F'_{\bar{n}n} & F'_{\bar{n}\bar{n}} \\
\end{pmatrix}
=
\begin{pmatrix}
	F_{nn} e^{-i\phi^n_s} & F_{n\bar{n}} \\
	 F_{\bar{n}n} & F_{\bar{n}\bar{n}} e^{i\phi^n_s} \\
\end{pmatrix} \nonumber \\
&=&
e^{-i\sigma_3\phi_s^n/2} \bm F_n e^{-i\sigma_3\phi_s^n/2}.
\end{eqnarray}
Then a spin Josephson supercurrent with spin at $n$ direction can be obtained.

(b) From the perspective of GL-type theory:
Considering the spin degree of freedom, the 1st GL-type equation near the critical point can be written as \cite{Bao2015_SSCGL1,Lv2017_SSCGL2}
\begin{equation}
	a{\bm \Psi}+\frac{1}{2m} [-i\hbar \partial_x -\alpha_0 ({\bm s} \times {\bm E})_x ]^2 {\bm \Psi} + {\bm s} \cdot {\bm B} {\bm \Psi}=0, \label{GL}
\end{equation}
where $ {\bm s=(s_x,s_y,s_z)} $ indicates the three $S=1$ spin operators,
and the $\alpha_0$ term represents SOC.
The equation's solution ${\bm \Psi= (\Psi^z_1, \Psi^z_0, \Psi^z_{-1})^T}$ represents  order parameters of $ S_z=1, 0, -1 $ components of spin-triplet Cooper pairs.
The initial system in Eq. (\ref{H_SNW}) has an $\bar x$-direction magnetic field and a $z$-direction $ \bm \alpha $, i.e. a $y$-direction electric field.
A $z$-direction spin phase effectively rotates the magnetic field, so that $ {\bm s} \cdot {\bm B} {\bm \Psi} $ is changed to
\begin{equation}
	\frac{1}{\sqrt{2}}
	\begin{pmatrix}
	0 & B e^{-i\phi_s^z} & 0 \\
	B e^{i\phi_s^z} & 0 & B e^{-i\phi_s^z} \\
	0 & B e^{i\phi_s^z} & 0 \\
\end{pmatrix}{\bm \Psi},
\end{equation}
while the first and second terms in Eq. (\ref{GL}) are diagonal and keep unchanged.
So the solution becomes $ \bm \Psi '= (\Psi^z_1 e^{-i\phi_s^z}, \Psi^z_0, \Psi^z_{-1} e^{i\phi_s^z})^T $.
This indicates that the $z$-direction spin phase adjusts the phases of spin-triplet $ S_z=1 $ and $ S_z=-1 $ order parameters by $-\phi_s^z$ and $ \phi_s^z $ respectively, thus the spin-$z$-direction spin supercurrent is caused.
As for an $x$-direction spin phase, in the $ (\Psi^x_1, \Psi^x_0, \Psi^x_{-1})^T $ basis, we can also find that only the kinetic term in Eq. (\ref{GL}) is changed to
\begin{equation} \frac{1}{2m}\begin{pmatrix}
	-i\hbar \partial_x & -\frac{i\alpha_0}{\sqrt{2}} E e^{-i\phi_s^x} & 0 \\
	\frac{i\alpha_0}{\sqrt{2}} E e^{i\phi_s^x} & -i\hbar \partial_x & -\frac{i\alpha_0}{\sqrt{2}} E e^{-i\phi_s^x} \\
	0 & \frac{i\alpha_0}{\sqrt{2}} E e^{i\phi_s^x} & -i\hbar \partial_x \\
\end{pmatrix}^2 {\bm \Psi}.
\end{equation}
Therefore, the triplet Cooper pairs of $S_x=1$ and $S_x=-1$ also get opposite phases and a spin supercurrent with spin pointing to $x$ direction is generated.

(c) Applying an $n$-direction spin phase $\phi_s^n$ is equivalent to rotating ${\bm d}$-vector by $\phi_s^n$ around direction $n$:
To show this we would like to prove that $\bm F'$, the pairing correlations with an $n$-direction spin phase, has a relation to the rotated vector $\bm d'$,
\begin{equation}
	\bm F'=(d_0+{\bm d'}\cdot{\bm \sigma}) i\sigma_2.\label{d}
\end{equation}
The $\bm F$ matrix in the original spin $z$ basis
and $\bm F_n$ in the spin $n$ basis are connected by $\bm F =U \bm F_n U^T$,
with $U$ the rotation operator in spin space
\begin{equation}
U=
\begin{pmatrix}
	\cos \frac{\theta}{2}  & -\sin \frac{\theta}{2} e^{-i\phi} \\
	 \sin \frac{\theta}{2} e^{i\phi}  & \cos \frac{\theta}{2}\\
\end{pmatrix} =
e^{-i\sigma_{n\perp}\theta/2}.
\end{equation}
Here $\sigma_{n\perp}=- \sigma_1 \sin \phi + \sigma_2 \cos\phi $ corresponds to
the Pauli matrix with the direction perpendicular to $nOz$ plane and
satisfies $ \sigma_2 \sigma^*_{n\perp} =-\sigma_{n\perp} \sigma_2 $.
It should be noted that the right operator is $U^T$ instead of $U^\dagger$.
This is because it operates on the hole space, and we should use the conjugate form of $U^\dagger$.
Relying on the relations above, we have
\begin{widetext}
	\begin{align}
		\bm F' &=e^{-i\sigma_{n\perp}\theta/2} \bm F'_n e^{-i\sigma_{n\perp}^* \theta/2}
		=e^{-i\sigma_{n\perp}\theta/2} e^{-i\sigma_3\phi_s^n/2} \bm F_n e^{-i\sigma_3\phi_s^n/2} e^{-i\sigma_{n\perp}^* \theta/2}\notag\\
		&=e^{-i\sigma_{n\perp}\theta/2} e^{-i\sigma_3\phi_s^n/2} e^{i\sigma_{n\perp}\theta/2} (d_0+{\bm d}\cdot{\bm \sigma})i\sigma_2 e^{i\sigma_{n\perp}^* \theta/2} e^{-i\sigma_3\phi_s^n/2} e^{-i\sigma_{n\perp}^* \theta/2} \notag \\
		&=e^{-i\sigma_{n\perp}\theta/2} e^{-i\sigma_3\phi_s^n/2} e^{i\sigma_{n\perp}\theta/2} (d_0+{\bm d}\cdot{\bm \sigma}) e^{-i\sigma_{n\perp} \theta/2} e^{i\sigma_3\phi_s^n/2} e^{i\sigma_{n\perp} \theta/2} i\sigma_2 \notag \\
		&=e^{-i \hat n \cdot {\bm \sigma} \phi_s^n/2} (d_0+{\bm d}\cdot{\bm \sigma}) e^{i \hat n \cdot {\bm \sigma} \phi_s^n/2} i\sigma_2. \label{relat}
	\end{align}
\end{widetext}
Note that in geometric algebra, the Pauli matrices behave as the basis vectors in real space, so that
\begin{equation}
	e^{-i \hat n \cdot {\bm \sigma} \phi_s^n/2} ({\bm d}\cdot{\bm \sigma}) e^{i \hat n \cdot {\bm \sigma} \phi_s^n/2}= {\bm d'}\cdot{\bm \sigma}. \label{rotate}
\end{equation}
Relying on Eq. (\ref{relat}) and Eq. (\ref{rotate}), Eq. (\ref{d}) is proved.
One can see that the effect of adding an $n$-direction spin phase is the same as
rotating $\bm d$-vector, and also induces a spin current, as is implied by Ouassou \emph{et al.} \cite{Linder2019_SJE8}.

(d) The spin phase ${\bm \phi}_s$ should not be treated as
three components of a classical vector:
In general, a spin phase can be written as ${\bm \phi}_s= (\phi_s^x,\phi_s^y,\phi_s^z)=\phi_s^n \hat n$
with $\phi_s^n =\sqrt{(\phi_s^x)^2+(\phi_s^y)^2+(\phi_s^z)^2}$.
To view the spin phase, it stands for opposite phases on $S=1$ Cooper pairs
with spins at the designated direction $n$ and $\bar n$.
One should not regard the three vector components as inducing phases $\phi_s^x,\phi_s^y,\phi_s^z$ on spin $x, y, z$ Cooper pairs respectively,
because Pauli matrices do not commute.

However, the three components are meaningful, and they represent the expectation values of phases for spin $x, y, z$ Cooper pairs.
For example, the wavefunction of $z$-spin Cooper pair can be expanded in $S_n$ basis,
$ | S=1, S_z=1 \rangle= \cos^2 \frac{\theta}{2} | S=1, S_n=1 \rangle - \frac{\sin \theta e^{i\phi}}{\sqrt{2}} | S=1, S_n=0 \rangle + \sin^2 \frac{\theta}{2} e^{2i\phi} | S=1, S_n=-1 \rangle. $
Because $S_n=1$ and $S_n=-1$ components get phases $-\phi_s^n$ and $\phi_s^n$ respectively, the expectation value of phase for $z$-spin Cooper pair is
$-\phi_s^n(\cos^4 \frac{\theta}{2}-\sin^4 \frac{\theta}{2})=-\phi_s^n \cos \theta =-\phi_s^z$.
At the same time, the $\bar z$-spin Cooper pair has an expectation value of phase $\phi_s^z$.

This is analogous to spin bias ${\bm V}_s=(V_s^x, V_s^y, V_s^z)=V_s^n \hat n$ with $V_s^n =\sqrt{(V_s^x)^2+(V_s^y)^2+(V_s^z)^2}$,
which generates a spin current by regulating occupations of electrons with opposite spins \cite{Sun2004_Spinbias}.
The spin bias ${\bm V}_s=V_s^n \hat n$ can be viewed as that
it changes the chemical potentials of $n$-spin and $\bar n$-spin electrons
by $eV_s^n$ and $-eV_s^n$, respectively.
We should not think of the components $V_s^x$, $V_s^y$, and $V_s^z$
as the splitting chemical potentials for $x$-spin and $\bar x$-spin electrons,
$y$-spin and $\bar y$-spin electrons, $z$-spin and $\bar z$-spin electrons,
separately.
Nevertheless, these components indeed reflect the expected
occupation numbers for the corresponding spin directions.

\section{\label{SEC4} Calculations}

Next we confirm the spin Josephson effect by calculating the current in a SNW-NNW-SNW junction, where the NNW locates at $0<x<d$.
The left SNW is described by Eq. (\ref{H_SNW}) with zero spin phase,
while the right SNW has a spin phase $\phi_s$ in $z$ or $x$ directions (BdG form of Eq. (\ref{H2}) or Eq. (\ref{H3})).
Inside the superconducting gap,
the current is carried by ABSs.
In this study, the current of a certain ABS is calculated by solving its wavefunction
and applying the definition of charge and spin currents in the NNW (see Appendix \ref{AA}).
The whole subgap current is contributed by all ABSs under Fermi surface $\epsilon=0$,
\begin{equation}
	I_{c(s)}=\sum_{\rm{ABS}} \theta(-\epsilon_{\rm{ABS}}) I_{c(s)}^{\rm{ABS}},
\end{equation}
where $ \theta $ is the unit step function.

Previous studies have shown that for charge Josephson effect, there is a simple relation between an ABS and its corresponding charge supercurrent: $I^{\rm ABS}_c=- \frac{2e}{\hbar} \frac{ \partial{\epsilon_{\rm{ABS}}} }{ \partial{\phi_c}}$ \cite{Bagwell1992_Jose,Zhang2013_Jose}.
For the spin Josephson effect, one may naturally guess that this relation is adjusted by replacing $2e$ by $\hbar$,
\begin{equation}
 I^{\rm ABS}_s=-\frac{ \partial{\epsilon_{\rm ABS}} }{ \partial{\phi_s} }. \label{Is}
\end{equation}
For our definition of spin phase, this relation is rigorously proved (see Appendix \ref{AB}).
Our results by using wavefunction of ABS and applying Eq. (\ref{Is}) are the same.
We emphasize that this is not just a coincidence of formulas,
but an indication that our whole study is self-consistent.

\begin{figure}
	\includegraphics[width=\columnwidth]{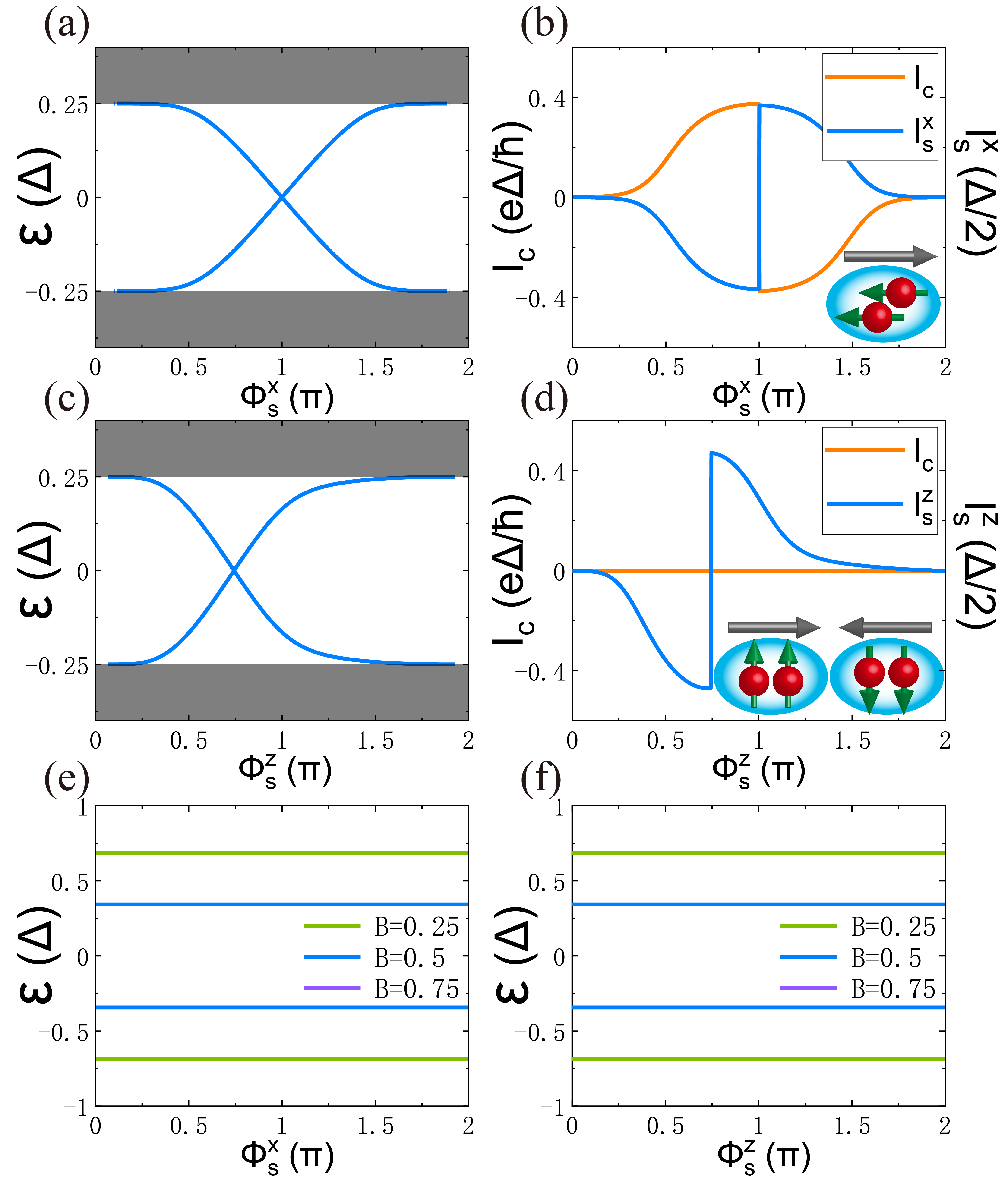}%
	\caption{\label{FIG3}
The ABSs (a, c) and supercurrent (b, d)
as functions of the $x$-direction (a, b) and $z$-direction (c, d)
spin phase difference with $B=1.25$ and NNW length $d \rightarrow 0$.
Inset in (b) shows the Cooper pairs almost polarized at $\bar x$-direction,
so in (b) the charge and spin supercurrents are nearly opposite with the ratio $-2e/\hbar$.
Inset in (d) shows that Cooper pairs can be equally decomposed into $z$ and $\bar z$ polarized components moving oppositely, and a pure spin supercurrent is realized in (d).
When SNWs are in trivial phase, the ABSs versus $x$ and $z$ direction spin phases are shown in (e, f) respectively.
The ABSs are independent of spin phase, so the subgap spin supercurrent is nonexistent according to Eq. (\ref{Is}).
For $B=0.75$, there is no ABS, so the corresponding curve is absent.
}
\end{figure}

Let's first consider the setup with an $x$-direction spin phase difference.
Here the SNWs on both sides are in topologically nontrivial phase ($B=1.25$).
We can see from Fig. \ref{FIG3}(a) that the energies of ABSs
are influenced by the spin phase,
thus a spin supercurrent emerges according to Eq. (\ref{Is}).
Because of the $\bar x$-direction magnetic field,
the $S=1$ Cooper pairs are mainly polarized at $\bar x$ direction.
Therefore, the supercurrent is caused by $\bar x$-spin Cooper pairs
and shows both charge current and $\bar x$-direction spin current in Fig. \ref{FIG3}(b).
This spin-polarized supercurrent reflects
the property that the SNW is both the charge superconductor and spin superconductor.

Next we study the setup with a $z$-direction spin phase difference.
In Fig. \ref{FIG3}(c), the ABSs are also regulated by the spin phase.
The dominated spin-triplet Cooper pairs are of spin-$\bar x$, of which the spin wavefunction can be expanded in the spin-$z$ basis: $ | S=1, S_x=-1 \rangle= \frac{1}{2} | S=1, S_z=1 \rangle - \frac{1}{\sqrt{2}} | S=1, S_z=0 \rangle + \frac{1}{2} | S=1, S_z=-1 \rangle $.
Therefore, the spin-$\bar x$ component is equally contributed by spin-$z$ and spin-$\bar z$ components.
Their opposite motions result in a pure $z$-direction spin supercurrent, while the charge current is canceled in Fig. \ref{FIG3}(d).

In addition, the subgap spin supercurrent
is a unique phenomenon of the topologically nontrivial phase,
because the gap originates from the coupling of electron
and hole bands with the same spin $\bar x$.
For the trivial phase ($ B<1 $),
the energy gap is formed by electron and hole with opposite spins.
Although there already exists spin-triplet components,
the equal-spin Andreev reflection coefficient is almost zero (see Fig. \ref{FIG1}(c)).
When the spin phases $\phi_s^x$ and $\phi_s^z$ are applied, the ABSs are not affected, as is shown in Fig. \ref{FIG3}(e,f).
When $B=0.75$, the ABS does not appear.
Because of the relation Eq. (\ref{Is}), the subgap spin current is almost zero in trivial phase.

\section{\label{SEC5} Discussion and conclusion}

The spin current we studied is carried
by $S=1$ Cooper pairs and can flow without dissipation.
Therefore, by non-local resistance
measurements \cite{Chen2013_Nonlocal,VanWees2015_Nonlocal,Han2018_Nonlocal},
this spin Josephson supercurrent can be detected far away from the junction interface.
The methods include using ferromagnetic electrodes \cite{Tinkham2006_Ferromagnetic,Chen2013_Nonlocal} and inverse spin Hall effect \cite{VanWees2015_Nonlocal,Shi2020_Inverse}.

In summary, we propose spin phase that generates spin Josephson supercurrent from spin-triplet superconductivity.
By controlling the spin current, magnetic field or electric field,
the spin phase, opposite phases on spin-up and spin-down Cooper pairs,
can be generated.
It leads to opposite motions of the spin opposite $S=1$ Cooper pairs
and gives rise to a spin supercurrent.
In the topologically nontrivial phase of SNW,
a subgap spin supercurrent can be induced by the spin phase,
while this subgap current is absent in the trivial phase.
We also derive a concise formula Eq. (\ref{Is}) about spin current, in analogy to charge Josephson effect.
Our findings offer a mechanism for spin Josephson effect, and put forward a spin supercurrent participated by topological phase transition.

\begin{acknowledgments}
    This work was financially supported by National Key R and D Program of China (Grant No. 2017YFA0303301), NSF-China (Grant No. 11921005),
	and the Strategic priority Research Program of Chinese Academy of Sciences (Grant No. DB28000000).
\end{acknowledgments}

\appendix
\begin{widetext}
\section{\label{AA} Calculation of charge and spin currents carried by ABSs}

In this Appendix, we show the method to derive the wavefunctions of ABSs and to calculate the corresponding charge and spin currents.

In the basis $(\psi_x, \psi_{\bar x}, \psi_x^\dagger, {\psi_{\bar x}}^\dagger)^T$, for an isolated NNW, the wavefunction at energy $\epsilon$ can be expanded by 4 electron modes and 4 hole modes,
\begin{equation}
	\psi_{\rm NNW} (x)= ( a_1 e^{i k_e x} +a_2 e^{-ik_e x}, a_3 e^{i k_e x} +a_4 e^{-ik_e x}  , b_1 e^{i k_h x} +b_2 e^{-ik_h x} , b_3 e^{i k_h x} +b_4 e^{-ik_h x} )^T,
\end{equation}
where $ k_e=\frac{1}{\hbar} \sqrt{2m(\mu_{\rm NNW}+\epsilon)} $ is the wave vector of electrons, $ k_h=\frac{1}{\hbar} \sqrt{2m(\mu_{\rm NNW}-\epsilon)} $ is the wave vector of holes,
and $a_{i}$, $b_{i}$ ($i=1,2,3,4$)
are coefficients for the electron and hole modes.

Let's take the setup in the presence of an $x$-direction
spin phase difference $\phi_s^x$ as an example.
The BdG form Hamiltonian of the right-side SNW is described by
\begin{equation}
	H_{\rm R}=\begin{pmatrix}
		\frac{p_x^2}{2m}+B & \alpha p_x e^{-i (\phi_s^x +\frac{\pi}{2})} & 0 & \Delta \\
		\alpha p_x e^{i (\phi_s^x +\frac{\pi}{2})} & \frac{p_x^2}{2m}-B & -\Delta & 0 \\
		0 & -\Delta & -\frac{p_x^2}{2m}-B & \alpha p_x e^{i (\phi_s^x +\frac{\pi}{2})} \\
		\Delta & 0 & \alpha p_x e^{-i (\phi_s^x +\frac{\pi}{2})} & -\frac{p_x^2}{2m}+B
	\end{pmatrix}.
\end{equation}

For an infinite long SNW, we can introduce the wave vector as a good quantum number $p_x=\hbar k$.
It has four energy bands,
\begin{equation}
	\epsilon_{\pm}^2=\xi^2+B^2+A^2+\Delta^2 \pm 2 \sqrt{\xi^2 A^2 + \xi^2 B^2 + B^2 \Delta^2}.
\end{equation}

Given an energy $ \epsilon $, the eigenvalue equation gives 8 solutions of $ k $.
These solutions represent the 8 modes in the SNW,
and each $ k_j $ can be traced back to one energy band $ \epsilon_l (k) $.
By a direct calculation, the wavefunction of the $j$th mode $ k_j $ is given by
\begin{equation}
	\psi_{{\rm R}, j} =\frac{1}{N_j} \begin{pmatrix}  \Delta \left[ (\epsilon_l+B)^2-\Delta^2-A^2-\xi^2 \right] \\
		 2\Delta A (B+\xi) e^{i (\phi_s^x +\frac{\pi}{2})}\\
		 A \left[ (\epsilon_l -T )^2-\Delta^2-A^2-B^2 \right] e^{i(\phi_s^x +\frac{\pi}{2})} \\
		  (\epsilon_l-\xi-B) \left[ (\epsilon_l +B)^2-\Delta^2+A^2-\xi^2 \right]-2\xi B^2  \end{pmatrix} e^{i k_j x},
\end{equation}
where $N_j$ is the normalization coefficient.
For the SNW on the left side, the 8 modes can be obtained by setting $ \phi_s^x=0 $.
Inside the superconducting gap, there is no real solution of $ k $.
The 4 modes for wave vectors with positive (negative) imaginary parts on the left (right) side
tend to infinity at $ x\rightarrow -\infty $ ($x\rightarrow+\infty$),
so these modes are not physically permitted.
The other 4 modes with negative (positive) imaginary parts
tend to 0 at $ x\rightarrow -\infty $ ($x\rightarrow+\infty$), which are physically permitted,
and they are marked as $\psi_{{\rm L}, i}$ ($\psi_{{\rm R}, i}$) with $i=1,2,3,4$.

Then the wavefunction of an ABS can be expressed as:
\begin{equation}
	\psi=\begin{cases}
		c_1 \psi_{{\rm L}, 1} + c_2 \psi_{{\rm L}, 2} + c_3 \psi_{{\rm L}, 3} + c_4 \psi_{{\rm L}, 4} & \text{if } x < 0,\\
		( a_1 e^{i k_e x} +a_2 e^{-ik_e x}, a_3 e^{i k_e x} +a_4 e^{-ik_e x}  , b_1 e^{i k_h x} +b_2 e^{-ik_h x} , b_3 e^{i k_h x} +b_4 e^{-ik_h x} )^T & \text{if } 0 < x < d,\\
		d_1 \psi_{{\rm R}, 1} + d_2 \psi_{{\rm R}, 2} + d_3 \psi_{{\rm R}, 3} + d_4 \psi_{{\rm R}, 4} & \text{if } x > d.
		\end{cases}
\end{equation}
This wavefunction should satisfy the continuity and current conservation
boundary conditions at $x=0$ and $x=d$ \cite{Sarma2015_Conductance},
\begin{equation}
	\psi|_{x=0-}=\psi|_{x=0+}, (\hat v \psi)|_{x=0-}=(\hat v \psi)|_{x=0+}, \psi|_{x=d-}=\psi|_{x=d+}, (\hat v \psi)|_{x=d-}=(\hat v \psi)|_{x=d+},
\end{equation}
where $ \hat v = \frac{1}{\hbar} \frac{\partial H}{\partial k}|_{k\rightarrow -i\partial_x} $ is the velocity operator.
Therefore, one gets a set of homogeneous linear equations of $X=(a_1, a_2, a_3, a_4, b_1, b_2, b_3, b_4, c_1, c_2, c_3, c_4, d_1, d_2, d_3, d_4)^T$.
The equations can be represented as $ SX=0 $, with $S$ the $16 \times 16$ transmission matrix.
In order to make the equations have solutions, the determination of $S$ must be zero, so that the ABSs can be solved only at specific energies.

In the NNW, the current of ABS is carried by electron and hole modes and keeps a constant.
But in the SNWs, the current decays to zero in our single particle representation, because it is converted to supercurrent carried by Cooper pairs.
So we should calculate the current in the NNW.
For a normalized ABS wavefunction, by applying the definition of charge current and $x$-direction spin current \cite{Sun2008_Definition}, they are found to be
\begin{equation}
	I_{c}^{\rm{ABS}}=e \frac{\hbar}{m} \left[k_e (|a_1|^2-|a_2|^2+|a_3|^2-|a_4|^2)+ k_h (|b_1|^2-|b_2|^2+|b_3|^2-|b_4|^2) \right] .\label{IC}
\end{equation}
\begin{equation}
	I_{s}^{{\rm{ABS}},x}=\frac{\hbar}{2} \frac{\hbar}{m} \left[k_e (|a_1|^2-|a_2|^2-|a_3|^2+|a_4|^2)+ k_h (|b_1|^2-|b_2|^2-|b_3|^2+|b_4|^2) \right] .\label{IS}
\end{equation}

\section{\label{AB} Derivation of the relation between ABS and spin current}

In this Appendix, we derive Eq. (\ref{Is}).
Let's also consider the case with an $x$-direction spin phase difference $\phi_s^x$.

Here we derive it in the basis $(\psi_x e^{i(\phi_s^x+\pi)/4}, \psi_{\bar x} e^{-i(\phi_s^x+\pi)/4}, \psi_x^\dagger e^{-i(\phi_s^x+\pi)/4}, {\psi_{\bar x}}^\dagger e^{i(\phi_s^x+\pi)/4})^T$.
It is different from the basis in Appendix \ref{AA} by only a unitary transformation in the whole space.
Because the expression of spin current Eq. (\ref{IS}) is not changed by this transformation, the equation $I^{\rm ABS}_s=-\frac{\partial{\epsilon_{\rm ABS}}}{\partial{\phi_s}}$ in this basis can be applied to the original basis.

The complete Hamiltonian of the Josephson junction is described by
\begin{equation}
	H = \left[\frac{p_x^2}{2m}-\mu(x) \right]\sigma_0 \tau_3 +\frac{1}{2} (\hat \alpha p_x + p_x \hat \alpha) + B(x) \sigma_3 \tau_3 -\Delta (x) \sigma_2 \tau_2,
\end{equation}
where $ \sigma $ and $\tau$ are Pauli matrices in spin and particle-hole spaces respectively, $ \mu(x)=\mu_{\rm NNW} \theta (x) \theta (d-x) $, $ B(x)=B \left[ \theta (-x) +\theta (x-d) \right] $, $ \Delta(x)=\Delta \left[ \theta (-x) +\theta (x-d) \right] $,
and $ \hat \alpha=\alpha \left[ \sigma_1 \tau_0 \cos \frac{\phi_s^x}{2}
+ \sigma_2 \tau_3 \sin \frac{\phi_s^x}{2} {\rm sgn} ( x-\frac{d}{2} ) \right]\left[ \theta (-x) +\theta (x-d) \right] $ is a
$4\times 4$ matrix.
Note that $ \hat \alpha $ is a function of coordinate $x$, and $ p_x $ is a differential operator on $x$.
Considering the Hamiltonian to be Hermitian, we write the SOC term as $ \frac{1}{2} (\hat \alpha p_x + p_x \hat \alpha) $ instead of $ \hat \alpha p_x $.

The wavefunction $\psi$ of ABS satisfies the Schr{\" o}dinger equation,
which can be expressed in the form
\begin{equation}
	\sigma_0 \tau_3 (\frac{p_x^2}{2m} \psi) + \hat \alpha (p_x \psi) +\frac{1}{2} (p_x \hat \alpha) \psi = \mu(x) \sigma_0 \tau_3 \psi - B(x) \sigma_3 \tau_3 \psi + \Delta (x) \sigma_2 \tau_2 \psi + \epsilon_{\rm ABS} \psi.\label{R1}
\end{equation}

By a transpose conjugate operation and taking the negative, we can write it as
\begin{equation}
	-(\frac{p_x^2}{2m} \psi^\dagger) \sigma_0 \tau_3  + (p_x \psi^\dagger) \hat \alpha +\frac{1}{2} \psi^\dagger (p_x \hat \alpha) = -\psi^\dagger \mu(x) \sigma_0 \tau_3 + \psi^\dagger B(x) \sigma_3 \tau_3 - \psi^\dagger \Delta (x) \sigma_2 \tau_2  - \psi^\dagger \epsilon_{\rm ABS}.\label{R2}
\end{equation}

We can apply $ \psi^\dagger \sigma_3 \tau_3 $ on the left of Eq. (\ref{R1}) and apply $\sigma_3 \tau_3 \psi $ on the right of Eq. (\ref{R2}), and then add them up.
This gives a relation
\begin{equation}
	\psi^\dagger \sigma_3 \tau_0 ( \frac{p_x^2}{2m} \psi )
		- ( \frac{p_x^2}{2m} \psi^\dagger ) \sigma_3 \tau_0 \psi
		+ \psi^\dagger \sigma_3 \tau_3 \hat \alpha ( p_x \psi )
		+ ( p_x \psi^\dagger ) \hat \alpha \sigma_3 \tau_3 \psi
		+ \frac{1}{2}\psi^\dagger \left[ ( p_x \hat \alpha ) \sigma_3 \tau_3+\sigma_3 \tau_3 ( p_x \hat \alpha ) \right] \psi=0.  \label{R3}
\end{equation}

To calculate the $x$-direction spin current,
we apply its definition in reference \cite{Sun2008_Definition},
$I^{{\rm ABS},x}_s = Re \left[\psi^\dagger \frac{1}{2}(\hat v \hat s+\hat s \hat v) \psi\right] = Re (J) $,
where $ \hat v= \frac{1}{\hbar} \frac{\partial H}{\partial k} |_{k\rightarrow-i\partial_x} $ is the velocity operator, $ \hat s = \frac{\hbar}{2} \sigma_3 \tau_3 $ is the spin operator.
The spin current $ I^{{\rm ABS},x}_s $ is the real part of $ J $, which has the form
\begin{equation}
	J=\frac{\hbar}{2} \psi^\dagger \left[ \frac{p_x}{m} \sigma_3 \tau_0 + \frac{1}{2}(\hat \alpha \sigma_3 \tau_3+ \sigma_3 \tau_3 \hat \alpha) \right] \psi.
\end{equation}

Applying $ \frac{p_x}{\hbar/2} $ on $J$, we can find a simple relation relying on Eq. (\ref{R3}),
\begin{equation}
	\begin{aligned}
		\frac{p_x}{\hbar/2} J
		={}& p_x (\psi^\dagger \frac{p_x}{m} \sigma_3 \tau_0 \psi)
		+ \frac{1}{2} p_x \left[\psi^\dagger (\hat \alpha \sigma_3 \tau_3
		+ \sigma_3 \tau_3 \hat \alpha) \psi\right] \\
	    ={}& \psi^\dagger \sigma_3 \tau_0 ( \frac{p_x^2}{2m} \psi )
		- ( \frac{p_x^2}{2m} \psi^\dagger ) \sigma_3 \tau_0 \psi
		+ \frac{p_x^2}{2m} ( \psi^\dagger \sigma_3 \tau_0 \psi )
		+ \frac{1}{2}( p_x \psi^\dagger ) (\hat \alpha \sigma_3 \tau_3
		+ \sigma_3 \tau_3 \hat \alpha) \psi \\
		&+ \frac{1}{2}\psi^\dagger (\hat \alpha \sigma_3 \tau_3
		+ \sigma_3 \tau_3 \hat \alpha) ( p_x \psi )
		+ \frac{1}{2}\psi^\dagger \left[( p_x \hat \alpha ) \sigma_3 \tau_3
		+ \sigma_3 \tau_3 ( p_x \hat \alpha )\right]\psi \notag\\
		={}& \frac{p_x^2}{2m} ( \psi^\dagger \sigma_3 \tau_0 \psi )
		-\frac{1}{2}( p_x \psi^\dagger )[\hat \alpha, \sigma_3 \tau_3] \psi
		+\frac{1}{2}\psi^\dagger [\hat \alpha, \sigma_3 \tau_3] ( p_x \psi )\\
		={}& \frac{p_x^2}{2m} ( \psi^\dagger \sigma_3 \tau_0 \psi )
		+ \psi^\dagger [\hat \alpha, \sigma_3 \tau_3] ( p_x \psi )
		+\frac{1}{2}\psi^\dagger [p_x \hat \alpha, \sigma_3 \tau_3] \psi
		-\frac{1}{2} p_x (\psi^\dagger [\hat \alpha, \sigma_3 \tau_3] \psi) .
	\end{aligned}
\end{equation}
By direct calculations, we can conclude that $ [\hat \alpha, \sigma_3 \tau_3]=-4i {\rm sgn} (x-\frac{d}{2}) \frac{ \partial \hat \alpha }{ \partial \phi_s^x } $, thus
\begin{equation}
	\begin{aligned}
		\frac{p_x}{\hbar/2} J
		={}& \frac{p_x^2}{2m} ( \psi^\dagger \sigma_3 \tau_0 \psi )
		-4i {\rm sgn} (x-\frac{d}{2}) \psi^\dagger \frac{\partial}{\partial \phi_s^x} (\frac{\hat \alpha p_x+p_x \hat \alpha}{2}) \psi
		-\frac{1}{2} p_x (\psi^\dagger [\hat \alpha, \sigma_3 \tau_3] \psi) \\
		={}& \frac{p_x^2}{2m} ( \psi^\dagger \sigma_3 \tau_0 \psi )
		-4i {\rm sgn} (x-\frac{d}{2}) \psi^\dagger \frac{ \partial H }{ \partial \phi_s^x } \psi
		-\frac{1}{2} p_x (\psi^\dagger [\hat \alpha, \sigma_3 \tau_3] \psi). \\
	\end{aligned}
\end{equation}

It is further transformed into a differential equation of $ J $ about the coordinate,
\begin{equation}
	{\rm sgn} ( x-\frac{d}{2} ) \left[\frac{d}{dx} J (x) + \frac{i\hbar^2}{4m} \frac{d^2}{dx^2} ( \psi^\dagger \sigma_3 \tau_0 \psi )+\frac{\hbar}{4} \frac{d}{dx} ( \psi^\dagger [\hat \alpha, \sigma_3 \tau_3] \psi )\right] = 2 \psi^\dagger \frac{ \partial H }{ \partial \phi_s^x } \psi .
\end{equation}

After an integral of the whole space, the first term on the left side becomes $ -2 J ( \frac{d}{2} ) + J (+\infty) + J (-\infty) $.
According to Appendix \ref{AA}, the ABS is composed of propagating modes in NNW and evanescent modes in SNWs, and the latter decays to zero at $ x= \pm \infty $.
So the first term is finally $ -2 J ( \frac{d}{2} ) $.
The second and third terms on the left side become $ \left[\frac{i\hbar^2}{4m} \frac{d}{dx} ( \psi^\dagger \sigma_3 \tau_0 \psi )+ \frac{\hbar}{4} \psi^\dagger [\hat \alpha, \sigma_3 \tau_3] \psi \right]( |^{+\infty}_{\frac{d}{2}} - |^{\frac{d}{2}}_{-\infty} ) $.
At $ x= \pm \infty $, $\psi$ is zero.
For $ 0<x<d $, $ \hat \alpha = 0 $.
Therefore, after the integral, 
the only existing term is $\left[-\frac{i\hbar^2}{2m} \frac{d}{dx} ( \psi^\dagger \sigma_3 \tau_0 \psi )\right](\frac{d}{2})$.
One can find that it has no real part and can be expressed as $2wi$, with $w$ a real number.
Using the Feynman-Hellman's theorem, the right side is $ 2\frac{\partial \epsilon_{\rm ABS}}{\partial \phi_s^x} $.
So we get $ J=wi-\frac{\partial \epsilon_{\rm ABS}}{\partial \phi_s^x} $.
At last, $ I^{{\rm ABS},x}_s = Re (J) =-\frac{\partial \epsilon_{\rm ABS}}{\partial \phi_s^x} $.

For the case with a $z$-direction spin phase difference, this relation can also be derived in a similar way.

\end{widetext}

\end{document}